\documentclass[fleqn,usenatbib]{mnras}

\usepackage{newtxtext,newtxmath}

\usepackage[T1]{fontenc}

\DeclareRobustCommand{\VAN}[3]{#2}
\let\VANthebibliography\thebibliography
\def\thebibliography{\DeclareRobustCommand{\VAN}[3]{##3}\VANthebibliography}

\usepackage{graphicx}	
\usepackage{amsmath}	

\title[Jet size scale in MAXI J1820+070 and V404 Cygni]{Probing the jet size of two Black hole X-ray Binaries in the hard state}

\author[S. Prabu et al.]{
S. Prabu,$^{1}$\thanks{E-mail: steveraj.prabu@curtin.edu.au}
J.C.A. Miller-Jones,$^{1}$
A. Bahramian,$^{1}$
C.M. Wood,$^{1}$
S.J. Tingay,$^{1}$
P. Atri,$^{2}$
\newauthor 
R.M. Plotkin,$^{3,4}$
and J. Strader,$^{5}$
\\
$^{1}$International Centre for Radio Astronomy Research, Curtin University, Bentley, WA 6102, Australia\\
$^{2}$ASTRON, Netherlands Institute for Radio Astronomy, Oude Hoogeveensedijk 4, 7991 PD Dwingeloo, The Netherlands\\
$^{3}$Department of Physics, University of Nevada, Reno, NV 89557, USA\\
$^{4}$Nevada Center for Astrophysics, University of Nevada, Las Vegas, NV 89154, USA\\
$^{5}$ Centre for Data Intensive and Time Domain Astronomy, Department of Physics and Astronomy, Michigan State University, East Lansing, MI 48824, USA
}

\date{Accepted XXX. Received YYY; in original form ZZZ}

\pubyear{2015}

\begin{document}
\label{firstpage}
\pagerange{\pageref{firstpage}--\pageref{lastpage}}
\maketitle

\begin{abstract}
Using multi-frequency Very Long Baseline Interferometer (VLBI) observations, we probe the jet size in the optically thick hard state jets of two black hole X-ray binary (BHXRB) systems, MAXI J1820+070 and V404 Cygni. Due to optical depth effects, the phase referenced VLBI core positions move along the jet axis of the BHXRB in a frequency dependent manner.  We use this ``core shift'' to constrain the physical size of the hard state jet. We place an upper limit of $0.3$\,au on the jet size measured between the 15 and 5 GHz emission regions of the jet in MAXI J1820+070, and an upper limit of $1.0$\,au between the $8.4$ and $4.8$\,GHz emission regions of V404 Cygni. Our limit on the jet size in MAXI J1820+070 observed in the low-hard state is a factor of $5$ smaller than the values previously observed in the high-luminosity hard state (using time lags between multi-frequency light curves), thus showing evidence of the BHXRB jet scaling in size with jet luminosity. We also investigate whether motion of the radio-emitting region along the jet axis could affect the measured VLBI parallaxes for the two systems, leading to a mild tension with the parallax measurements of Gaia. Having mitigated the impact of any motion along the jet axis in the measured astrometry, we find the previous VLBI parallax measurements of MAXI J1820+070 and V404 Cygni to be unaffected by jet motion. With a total time baseline of $8$ years, due to having incorporated fourteen new epochs in addition to the previously published ones, our updated parallax measurement of V404 Cygni is $0.450 \pm 0.018$\,mas ($2.226 \pm 0.091$\,kpc).
\end{abstract}




\section{Introduction}
One of the many ways in which the gravitational potential energy of matter accreting onto a black hole is liberated is via the formation of bipolar relativistic jets \citep{2016LNP...905...65F}. Whilst these jets have been observed in many active galactic nuclei (AGN) in the past, they were not confidently accepted to be part of accreting stellar mass black hole systems until the 2000s \citep{2006csxs.book..381F}. These accreting stellar mass black holes, known as black hole X-ray binaries (BHXRBs), exist in different characteristic accretion states,  broadly classified into hard and soft states. In the hard state, we expect to see a power law X-ray spectrum thought to originate primarily from thermal comptonisation in the corona \citep{2006ARA&A..44...49R}, and a weak, partially self-absorbed compact jet \citep{2014SSRv..183..323F}. The hard state jet  displays a flat or slightly inverted ($f_{\nu} \propto \nu^{\alpha}$ radio spectrum, where $f_{\nu}$ is the flux density at an observing frequency ${\nu}$,  and $\alpha \approx 0$), which is interpreted as a superposition of multiple synchrotron components originating from different regions along the partially optically thick jet \citep{1979ApJ...232...34B}. During an outburst, the X-ray spectrum moves to a softer spectrum with a multi-temperature blackbody profile, due to thermal emission from the hot inner accretion disk \citep{2006ARA&A..44...49R}. During the hard-to-soft state transitions, BHXRBs often eject discrete knots of radio emitting material that can be seen as optically thin synchrotron emission. \\

The brightest radio-emitting region in the hard state jet is the surface at an optical depth of unity ($\tau \sim 1$) and its apparent position moves downstream with decreasing observing frequency. In this work, we aim to use this observed shift in core position (the so-called core shift) to probe the physical size of the hard state jet. For a given observing frequency the distance ($\Delta r$) from the black hole to the hard state jet photosphere varies as $\Delta r \propto \nu^{-1/k}$ \citep{1979ApJ...232...34B}. For a conical jet geometry in equipartition, $k=1$, which is shown to be appropriate for the majority of AGN \citep{2009MNRAS.400...26O,2011IAUS..275..194F,2012Natur.489..326H}.\\

\begin{figure}
\begin{center}
\includegraphics[width=\linewidth,keepaspectratio]{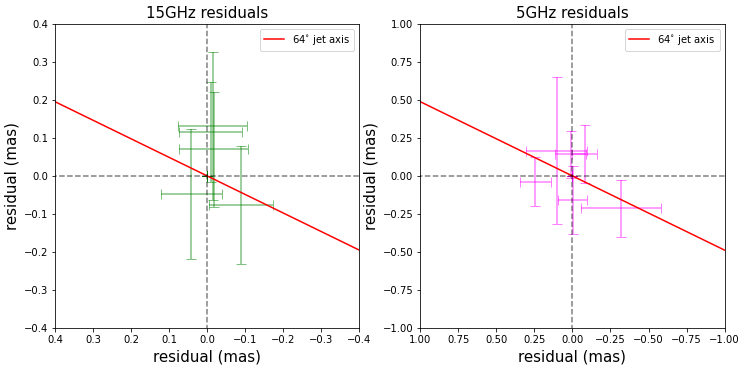}
\caption{2D astrometry fit residuals for $15$\,GHz and $5$GHz VLBI observations of MAXI J1820+070. Note that the red line is the jet axis of the system and is not fit to the data. }
\label{Fig1820residuals}
\end{center}
\end{figure}

\begin{figure}
\begin{center}
\includegraphics[width=\linewidth,keepaspectratio]{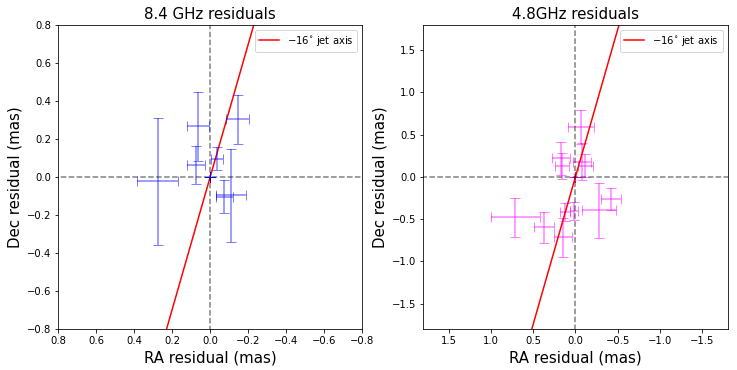}
\caption{2D astrometry fit residuals for $8.4$\,GHz and $4.8$\,GHz VLBI observations of V404 Cygni. Note that the red line is the jet axis of the system and is not fit to the data.}
\label{FigV404residuals}
\end{center}
\end{figure}

A tell-tale signature of the radio observation of the BHXRB being affected by optical depth effects is when the residuals of the astrometric fit are scattered in the direction of the jet axis of the system, as previously observed for Cygnus X-1 \citep{2012MNRAS.419.3194R}. In this paper, we investigate if the scatter in astrometry residuals of MAXI J1820+070 and V404 Cygni (Figure \ref{Fig1820residuals} and Figure \ref{FigV404residuals}, and the corresponding astrometry modeling is discussed in Section \ref{sec1D}) indicate any potential impact on their measured parallax, whilst also probing their jet scales measured between different frequencies. \\

Our motivation to use V404 Cygni as one of the objects of interest in this work is two-fold. First, we aim to probe the physical size of V404 Cygni's hard state jet as discussed above. Second, there exists a mild tension between V404 Cygni's radio parallax measurement of $0.418 \pm 0.024$\,mas using  Very Long Baseline Interferometery (VLBI) and its optical parallax measurement of $0.347 \pm 0.078$\,mas by the Gaia satellite \citep{1995ESASP.379..109T,2016A&A...595A...1G, 2018ApJS..239...31B} after correcting for the zero-point offset \citep{2021A&A...654A..20G}. There are two ways to measure the parallax signature of BHXRBs. As the blackbody spectrum of most stars peaks in the optical wavelengths, the donor star's parallax can be measured using optical telescopes like Gaia. Radio parallax of BHXRBs can also be measured by observing the powerful radio outflows from the accreting black holes using VLBI. Ideally, the Gaia and VLBI parallax measurements should be in agreement, but recently the tension in parallax measurements of Cygnus X-1 (Cyg X-1) was attributed to the unaccounted motion of the radio emission centroid along the jet axis of the BHXRB \citep{2021Sci...371.1046M}. \cite{2019ApJ...874...13P} and \cite{2022MNRAS.516.4640D} have found evidence of radio flares in hard state BHXRBs that change the flux density of the quiescent jet (by factors of 2 - 4), thus warranting the investigation of previously derived radio astrometry of the BHXRBs.\\\

\subsection{BACKGROUND}
\label{sec:background}

\subsubsection{MAXI J1820+070 (ASASSN-18ey)}
MAXI J1820+070 was discovered as an optical and X-ray transient by the All-Sky Automated Survey for SuperNovae \cite[ASAS-SN,] []{2018ApJ...867L...9T} and the Monitor of All-Sky X-ray Image \citep[MAXI,][]{2018ATel11399....1K}, respectively, as it went into outburst in 2018. \cite{2019ApJ...882L..21T} dynamically confirmed the system to host a black hole of mass $7-8M_{\odot}$ in a binary orbit with a K-type donor star. As MAXI J1820+070 transitioned from the hard to soft state during the outburst, \cite{2020NatAs...4..697B} and \cite{2021MNRAS.505.3393W} detected multiple bi-polar ejections along a jet axis of $64^{\circ} \pm 5 ^{\circ}$(measured East of North), which we adopt as the position angle of the jet in this work. Ejecta were also seen at a consistent position angle in X-ray during the outburst \citep{2020ApJ...895L..31E}. \cite{2020MNRAS.493L..81A} observed MAXI J1820+070 during the hard state of the rising outburst in 2018 and again as it faded to quiescence in February 2019, and measured its radio parallax to be $0.348 \pm 0.033$ mas. The optical zero-point corrected \citep{2021A&A...649A...4L} parallax measurement of MAXI J1820+070 by Gaia Data Release 3 \citep[DR3]{gaiacollaboration2022gaia} is  $ 0.398 \pm 0.078$ \,mas.\\

\subsubsection{V404 Cygni}
V404 Cygni has a dynamically confirmed black hole orbiting a K-type donor star \citep{1992Natur.355..614C, 1994MNRAS.271L..10S}. The black hole's mass is estimated to be $9.0^{+0.2}_{-0.6} M_{\odot}$ \citep{ 2010ApJ...716.1105K} and the quiescent radio luminosity of the system is particularly high due to its long orbital period. Its radio parallax was measured as $0.418 \pm 0.024$ mas \citep{2009ApJ...706L.230M} and its optical zero-point corrected parallax as reported in Gaia DR3 is $0.347 \pm 0.078$ mas.  During its outburst in 1989, a polarisation study performed by \cite{1992ApJ...400..304H} measured its electric vector position angle (EVPA) at multiple frequencies to be $-16^{\circ} \pm 6^{\circ}$ (measured East of North), later identified by \cite{2019Natur.569..374M} to be consistent with the jet direction. During its 2015 outburst the jet axis was observed to vary between $-30.6^{\circ}\pm 0.9^{\circ}$ and $+5.4^{\circ}\pm0.8^{\circ}$ \citep{2019Natur.569..374M}, which was attributed to Lense-Thirring precession of the inner (puffed up) accretion disk. However, in this work we use $-16^{\circ} \pm 6^{\circ}$ as the position angle of the jet axis, as the precession of the jets is only expected to occur when the source is accreting at super Eddington rates (not the case for the hard state and quiescent observations considered here). \\

\begin{figure*}
\begin{center}
\includegraphics[width=0.8\linewidth,keepaspectratio]{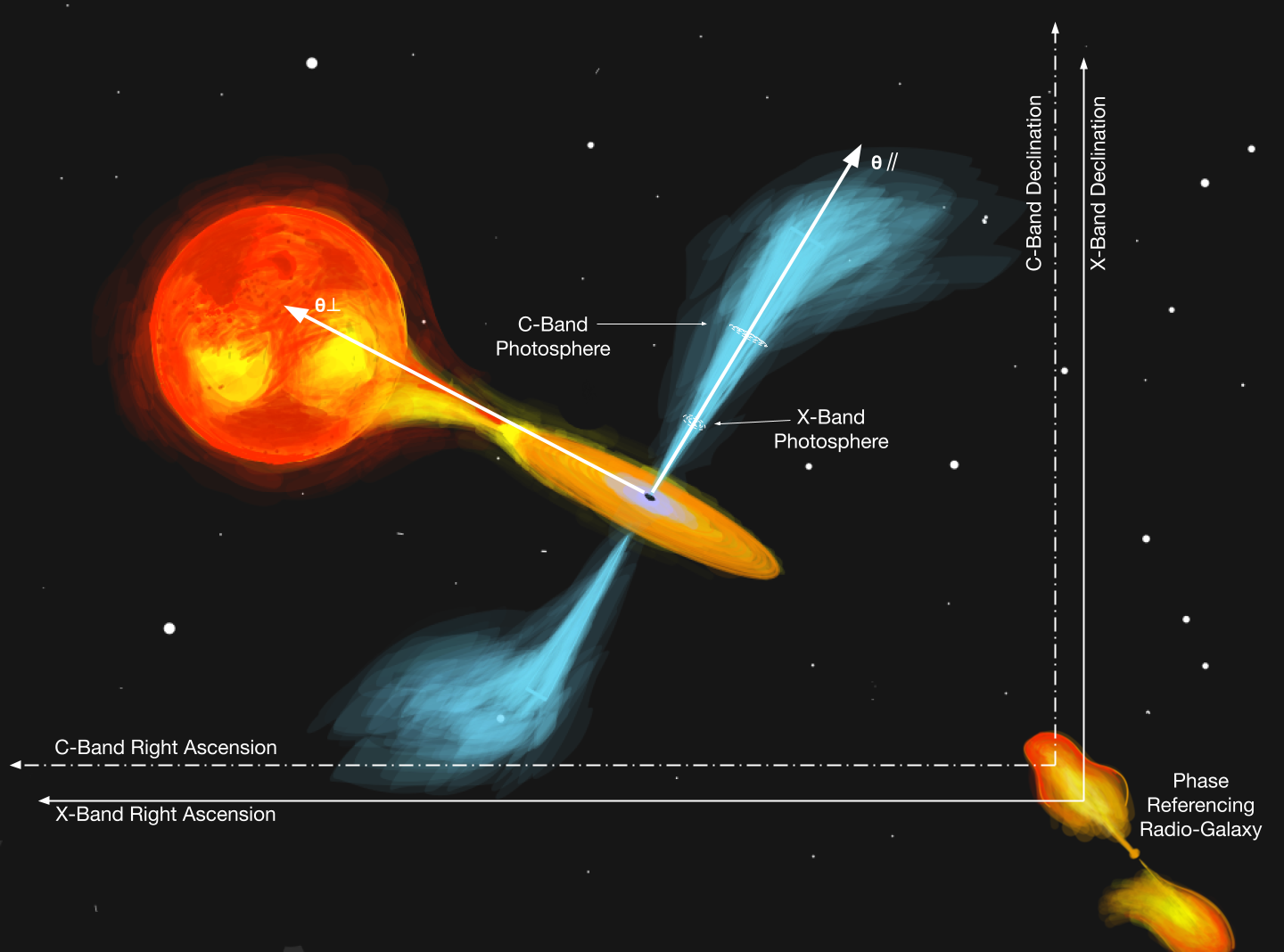}
\caption{An artistic impression of a BHXRB accreting through Roche-lobe overflow in the hard state and the corresponding astrometric coordinate systems considered in this work. The position of the BHXRB is measured with respect to a phase referencing source (radio galaxy in the bottom-right of the Figure), and we also show the core shift contribution from the phase referencing source when observing at different frequencies. The figure also shows the 1D astrometric coordinates used in this work denoted by $\theta_{\perp}$ and ${\theta_{\parallel}}$. NOTE: the $\theta_{\perp}$ and ${\theta_{\parallel}}$ coordinate system is still centred on the phase reference source. We only offset it to the position of the black hole in the figure for better illustration.}
\label{Fig1coreshift}
\end{center}
\end{figure*}

\subsubsection{Jet size studies}
Many of the techniques used to measure the sizes of optically thick jets from accreting black holes originate from VLBI studies of AGN. Using the $\Delta r \propto \nu^{-1/k}$ model for core shift \citep{1979ApJ...232...34B}, \cite{2012Natur.489..326H} inferred the location of the central black hole in the radio galaxy M87. \cite{2012A&A...545A.113P} studied the core shift in $191$ AGNs (at a wide range of red-shifts) as part of the Monitoring of Jets in Active Galactic Nuclei with VLBA Experiments (MOJAVE) survey and placed parsec-scale size limits on the optically thick AGN jets. An alternative method of estimating jet sizes was developed by \cite{2011MNRAS.415.1631K} that used the time lags of flares observed at different frequencies to measure the jet size. The lag correlation method assumes a standard shock-in-a jet model where any change in pressure at the base of the conical jet propagates as a shock downstream, and hence emissions at longer wavelengths are a delayed version of the emission originating upstream much closer to the jet base \citep{2011MNRAS.415.1631K}. If the observed delay between two frequencies is $\Delta t$, the inclination of the jet with respect to the line of sight $\theta$, and jet speed $\beta$ (where velocity=$\beta c$), the distance between the emission regions is given by $\Delta z = \beta c \Delta t (1 - \beta cos \theta)^{-1}$ \citep{2019MNRAS.484.2987T}. However, a limitation of the delay correlation method is that it involves either having prior knowledge of the jet speed and inclination angle, or performing a joint fit for jet speed, inclination angle, and core shift. \cite{2019MNRAS.484.2987T} extended this technique to perform a multi-frequency lag correlation of high-hard\footnote{before the peak of the outburst.} state observations of MAXI J1820+070.\\

The hard state jets in BHXRBs have been observed to be resolved in a few systems: Cygnus X-1 \citep{stirling1998high}; GRS 1915+105 \citep{2000ApJ...543..373D}; MAXI J1836-194 \citep{2015MNRAS.450.1745R}; and MAXI J1820+070 \citep{2021MNRAS.504.3862T}. But only \cite{2012MNRAS.419.3194R} have used VLBI astrometry residuals to place limits on the jet size of a BHXRB (Cygnus X-1). \cite{2021MNRAS.504.3862T} performed a simultaneous multi-wavelength fast timing study of MAXI J1820+070 in the high hard state to measure the jet size, and we later compare their measurements to our values in Section \ref{sec:discussion}. \cite{plotkin20172015} and \cite{2019MNRAS.484.2987T} measured time lags between the X-ray and radio lightcurves for V404 Cygni and Cygnus X-1, respectively, and placed limits on the distance of the radio-emitting photospheres from their black holes. The X-ray emission originates from the inner accretion disk very close to the black hole and the radio emission originates from the jet, and the two light curves show a correlation due to disk-jet coupling. X-ray variations in the inner disk are assumed to propagate into the jet and flow downstream, where we observe them at the optical depth unity surface, and the measured delay between X-ray and radio lightcurves can be used to constrain the distance of the radio emission from the black hole. \cite{plotkin20172015} further verified their time-lag measurement of the jet in V404 Cygni by comparing it to the angular size upper limit from their Very Large Baseline Array (VLBA) observations of the unresolved source.\\

In this study, we fit for the core shift between the different frequencies, whilst also performing parallax and proper motion measurements using BHXRB astrometry measured perpendicular to its jet axis (following \cite{2021Sci...371.1046M}). This helps us probe the hard state jet scale in the two BHXRBs whilst also investigating the possibility that previous VLBI astrometry was affected by scatter along the jet axis. \\

This paper is structured as follows. In Section \ref{sec:observationsandmethods} we discuss the VLBI observations used and the techniques employed to perform astrometry for the two BHXRBs. We provide our results in Section \ref{sec:results}. We discuss our results and their implications in Section \ref{sec:discussion}. The paper is summarised in Section \ref{sec:conclustion}.\\

\section{Observations and Methods}
\label{sec:observationsandmethods}
In this work we use a combination of previously-processed VLBI observations and new data to perform astrometry. We measure both the BHXRB's parallax and proper motion in the sky relative to a nearby extragalactic background source (with an assumed J2000 ICRF position obtained from the Radio Fundamental Catalogue\footnote{\url{http://astrogeo.org/vlbi/solutions/rfc_2015a/rfc_2015a_cat.html}}) using phase-referencing \citep{zensus1995very}. \\

\subsection{ASTROMETRY DATA AND PROCESSING}
\subsubsection{MAXI J1820+070}
For MAXI J1820+070, we use the $11$ radio measurements obtained by \cite{2020MNRAS.493L..81A} (also provided in Table \ref{tab:J1820}), to which we direct the reader for details on data processing. MAXI J1820+070 was observed at $15$\,GHz using the Very Long Baseline Array (VLBA) and at $5$\, GHz by the European VLBI Network (EVN) during the rising hard state of its 2018 outburst, and again after the peak before it entered the soft state. For the later epochs, the VLBA observations were performed at $5$\,GHz as the source faded. The last two VLBA observations were performed at both $5$\,GHz and $15$\,GHz in order to tie together the reference frames. When combining the VLBI phase-referenced observations of different frequencies, there are two different systematic offsets that can affect the data. First, we have the offset due to having two different phase-references sources (in phase-referencing experiments the assumed ICRF position of the phase reference source during correlation becomes the origin from which we measure the target's astrometry), and second, we have the frequency dependant core-shift affecting astrometry measurements made at different frequencies. \cite{2020MNRAS.493L..81A} calculated the systematic offset due to having used two different phase-reference sources to be $-0.29\pm0.08$ and $-0.05\pm0.02$ mas in RA and Dec directions, respectively, and accounted for the frequency dependant offset in their astrometry fitting code of MAXI J1820+070.\\

\subsubsection{V404 Cygni}
Unlike MAXI J1820+070, which is only observable by VLBI during the hard state immediately before/after an outburst, 
V404 Cygni is much brighter (and also intrinsically much more luminous) in the hard state where the systems spend most of their time, and hence we have many more observations of the source (Table \ref{tab:v404}). We use a total of 19 epochs (spread across $8$ years) to constrain V404 Cygni's astrometry, of which 5 epochs were obtained from \cite{2009ApJ...706L.230M}, and the remainder analysed here for the first time. We use a combination of $8.4$\,GHz and $4.8$\,GHz phase-referenced archival observations of the source, observed during different VLBI campaigns with different science goals.\\

The two new $8.4$\,GHz observations (epochs V06 and V19 in Table \ref{tab:v404}) were calibrated and imaged in the Astronomical Image Processing System \citep[{\tt AIPS};] [version 31 DEC22]{Greisen2003}, following the standard astrometry recipe described in the {\tt AIPS} cookbook\footnote{\url{http://www.aips.nrao.edu/cook.html}}. In addition, we remove clock errors and residual tropospheric effects from epoch V19 by running the {\tt AIPS} task {\tt DELZN} \citep{mioduszewski2009strategy} on geodetic blocks. All V404 Cygni target observations were phase referenced to the same extragalactic background source (J2025+3343\footnote{Note that we used VLBA observations of the target from four different campaigns, all of which assumed the same position of phase-reference source.}) and we verified our data processing by applying the calibration solutions to a nearby check source (J2023+3153, a different background source $1.87^{\circ}$ away). The remaining $8.4$\,GHz observations of V404 Cygni did not have geodetic blocks, but due to the very small angular separation ($16$ arcminutes) between the target and phase-reference source, \cite{2009ApJ...706L.230M} estimate the systematics to be on the order of $30\mu$as, which were added in quadrature to the measured source position errors.\\

 The twelve $4.8$\,GHz observations of V404 Cygni (epochs from V07 to V18 in Table \ref{tab:v404}) did not have geodetic observation blocks, and were more prone to systematic errors. Low elevation scans view the source through multiple different columns of the ionosphere and can leave behind residual delays despite having run the {\tt VLBATECR} task in {\tt AIPS}. The ionosphere can also become turbulent during sunrise/sunset and hence can leave behind unmodeled residual delays in the data, especially at lower frequencies. Hence, we perform rigorous flagging of the data to mitigate systematic errors as much as possible. All low elevation scans ($<23^{\circ}$)\footnote{as suggested by Mark Ried to James C.A. Miller-Jones through personal communication to be a good rule of thumb to mitigate ionospheric effects.} and scans 1hr before/after sunset/sunrise at each station were flagged. \\

The $4.8$\,GHz observations were also $90$ minutes in duration and had sparse uv-coverage on the phase reference source, which could manifest as larger systematics on V404 Cygni's astrometry. As the phase-reference source is resolved (due to scatter broadening), different epochs could pick different centroid positions of the resolved source as the origin of the reference frame of our relative astrometry. We mitigate this by making a 12 epoch stacked (using {\tt AIPS} task {\tt DBCON}) global model of the phase reference source, and by phase referencing the individual observations of V404 Cygni to the global model. Stacking the different epochs provides the best possible uv-coverage, and applies a consistent calibrator model to each epoch when fringe fitting.  The $12$ epochs were performed as filler-time observations and together spanned $6$ months. It is reasonable to assume (to first order) that the $4.8$\,GHz structure of the phase reference source does not evolve during the 6 months of observation. Any systematics that may arise from this are considered and discussed in Section \ref{sec:results}. \\

The global model was phase-only self-calibrated and imaged with {\tt Difmap} \citep{1997ASPC..125...77S}, and then read back into {\tt AIPS} for fringe fitting. Baselines longer than 90 mega-wavelengths in the global model were flagged as they appeared to be affected by scatter broadening\footnote{the light from the source is scattered by the interstellar medium along the propagation path and hence appears to be resolved in the long baselines. Using these long baselines can cause degradation in the resolution of astrometry measurements, and hence where flagged.}. The individual epochs were also inspected by eye to make sure that the flux density scales were aligned before stacking. Note we do not perform any amplitude self-calibration on the global model as any small changes in source flux density between the 12 epochs could leave behind spurious structures in the global model. \\

\subsection{ASTROMETRIC MODELLING}
\label{sec1D}
Astrometric fitting in the literature often uses the mathematical formalisation given in \cite{2007ApJ...671..546L}, to determine the parallax ($\pi$), proper motion ($\mu_{\alpha}\cos\delta$, $\mu_{\delta}$), and reference position (${\rm RA}_{0}, {\rm Dec}_{0} $) from multi-epoch RA-Dec measurements of the source. However, as we aim to investigate the presence of any resolved jet motion in our targets, we adopt a 1D astrometry technique developed by \cite{2021Sci...371.1046M} and shown in Figure \ref{Fig1coreshift}. We rotate our RA-Dec source position by the jet axis angle ($\theta$) into a new orthogonal reference frame measured perpendicular ($\theta_{\perp}$) and parallel ($\theta_{\parallel}$) (see Figure \ref{Fig1coreshift}) to the jet axis, and only use the $\theta_{\perp}$ component for astrometry as it would not be affected by any potential jet motion.\\

We perform the fitting using a {\tt PYMC3} \citep{salvatier2016probabilistic} implementation of the Hamiltonian Monte-Carlo \cite[HMC][]{2011hmcm.book..113N} technique. For modeling the motion of the BHXRBs, we use the modeling framework provided in \cite{2021Sci...371.1046M}. Note that we do not consider the orbital motion of the black hole for either of the systems studied here, as it is smaller ($3.8 \times 10^{-3}$mas for V404 Cygni and $6.7 \times 10^{-4}$mas for MAXI J1820+070) than the precision of the observations used here (see Appendix \ref{sec:orbitalmotion}). \\

When combining target positions measured at different frequencies, we expect core shifts to affect our astrometry. As previously mentioned, at correlation, we assume a particular position for the phase-reference source (the same source position is used for the different frequencies), and the observed radio emission of the background quasar is implicitly set at the assumed position during calibration. However, the observed emission  of the background quasar comes from different locations along the optically thick jet, thus shifting the coordinate system along its jet axis. Much like the quasar, the BHXRB also has an optically thick compact jet in the hard state and could have a core shift contribution affecting the astrometry if not accounted for. Hence, the core shift fit in our astrometry can be used to probe the jet scale of the BHXRB, and we discuss how the core shifts from the phase-reference source could affect our experiment. \\

We perform our astrometry in four steps:
\begin{enumerate}
    \item model the 2D astrometric positions (RA-Dec measurements) over time with $\pi$, $\mu_{\alpha} \cos\delta$, $\mu_{\delta}$, ${\rm RA}_{0}$, ${\rm Dec}_{0}$, $\Delta \delta$, and $\Delta \alpha$ (core shifts in RA and Dec directions) as model parameters;
    \item add the inferred core shifts (mean of the posteriors) to the RA-Dec measurements of our targets, and add the uncertainties in the core shift in quadrature to the existing errors in RA-Dec;
    \item rotate our updated RA-Dec (by the jet angle of the BHXRB) measurements from the previous step into $\theta_{\perp}$ and $\theta_{\parallel}$ directions; and
    \item model the 1D astrometric position ($\theta_{\perp}$) with $\pi$, $\mu_{\alpha}\cos\delta$, $\mu_{\delta}$,$\rm RA_{0}$, and $\rm Dec_{0}$ as model parameters. Due to degeneracy in only considering one axis of the source position, we use the mean and standard deviation of the posteriors from step (i) as priors for the 1D fit (for all the five parameters being fit for in this step). The parallax and proper motion determined in this step should not be influenced by any potential scatter along the jet axis in the BHXRB.
\end{enumerate} 

\subsection{CORE SHIFT TO JET SIZE}
\label{sec:jetsize}
We place upper limits on the jet size of the BHXRBs through boot-strap re-sampling \citep{dixon2006bootstrap} of the trace values of core shift and parallax obtained from our astrometry modelling. The sequence of steps in the boot-strap re-sampling is given below

\begin{enumerate}
    \item randomly sample a core shift in RA-Dec directions and a parallax value;
    \item rotate the sampled core shift to $\beta_{\perp}$ and $\beta_{\parallel}$ directions (where $\beta$ is the jet axis angle);
    \item calculate the jet size in the plane of the sky using parallax distance and $\beta_{\parallel}$ component of the core shift;
    \item using the known inclination angle of the jet, re-project the jet size projected on the sky plane in the direction of the jet axis.
    
\end{enumerate}

\section{Results}
\label{sec:results}

\begin{figure*}
\begin{center}
\includegraphics[width=\linewidth,keepaspectratio]{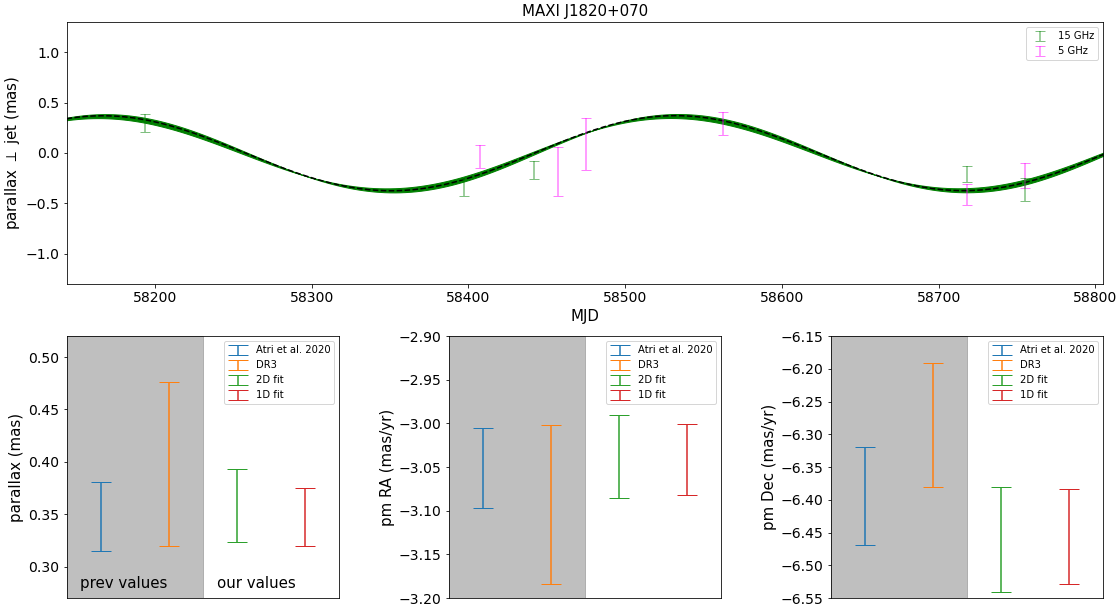}
\caption{Top panel shows the parallax and the source positions in the $\theta_{\perp}$ direction as a function of MJD. The black line is the mean of the 1D astrometry and the green shaded area shows the $68\%$ highest density interval. The bottom three panels show the parallax and the proper motion of the source determined during the 2D fit and 1D fit along with previous Gaia Data Release 3 (DR 3) and VLBI measurements. Our final 1D fit parallax and proper motion measurements for MAXI J1820+070 are also given in Table \ref{tab:results}.}
\label{fig1820results}
\end{center}
\end{figure*}

\begin{figure*}
\begin{center}
\includegraphics[width=\linewidth,keepaspectratio]{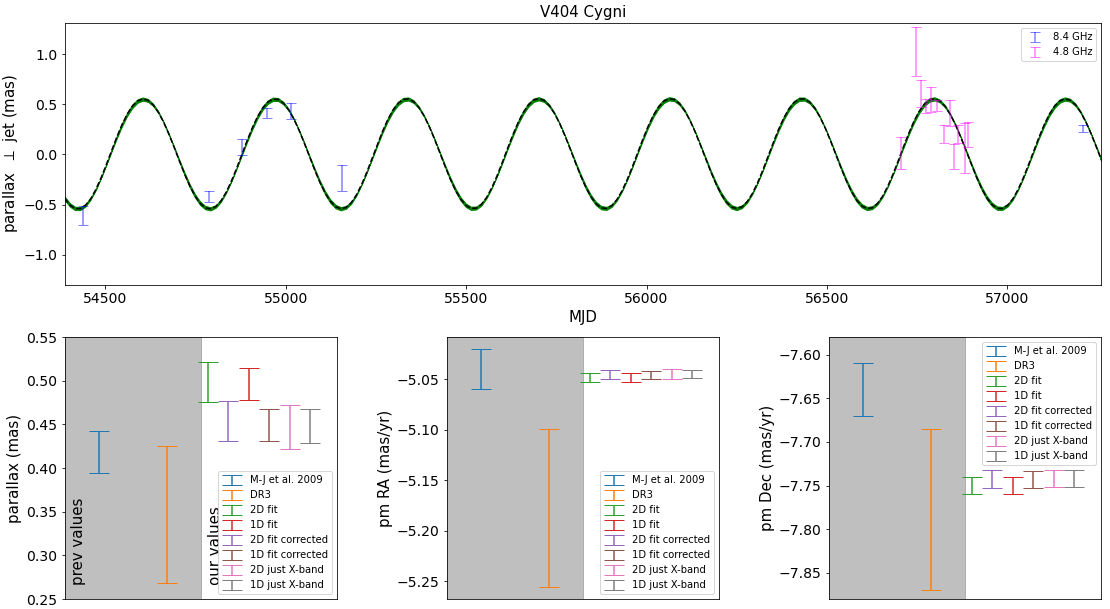}
\caption{ Top panel shows the parallax and the source positions in the $\theta_{\perp}$ direction as a function of MJD. The black line is the mean of the 1D astrometry and the green shaded area shows the $68\%$ highest density interval. The bottom three panels show the parallax and the proper motion of the source determined during the 2D fit and 1D fit along with previous Gaia Data Release 3 (DR 3) and VLBI measurements. Our final 1D fit parallax and proper motion measurements for V404 Cygni are also given in Table \ref{tab:results}. Note that in Section \ref{sec:results} we identify a systematic error affecting our $5$\,GHz observations of V404 Cygni. Hence, we re-do the 2D and 1D fit having corrected for the systematic, and show the corresponding systematic corrected parallax and proper motion in the bottom three plots (using purple and brown markers). To build confidence on our astrometry, we again do the fitting just using the X-band ($8.4$\,GHz) observations (as it was less prone to systematics) and obtain solutions that are in strong agreement with our previous C-band ($4.8$\,GHz) and X-Band joint fit. Our final 1D fit parallax and proper motion measurements for V404 Cygni, unaffected by jet motion, are also listed in Table \ref{tab:results}.}
\label{figV404results}
\end{center}
\end{figure*}

\begin{figure*}
\begin{center}
\includegraphics[width=0.7\linewidth,keepaspectratio]{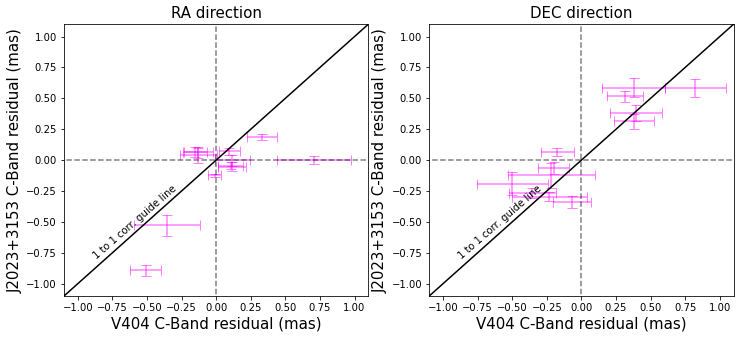}
\caption{Figure shows the observed approximate one-to-one correlation between C-band residuals of V404 Cygni with the residuals of the corresponding C-band check source.}
\label{FigV404VsChecksource}
\end{center}
\end{figure*}

The Gelman Rubin metric \citep[$\hat{R}$][]{gelman1992inference} is often used to evaluate the convergence of parallel HMC chains. The method compares the variance between the chains to the variance within the chain to determine if the simulation has converged. We obtain a value of $ \hat{R} <1.001$ for our astrometry modeling of both BHXRBs, well within the recommended $\hat{R} \leq 1.05$\footnote{\url{https://mc-stan.org/rstan/reference/Rhat.html}}. The astrometric residuals of the two BHXRBs are shown in Figure \ref{Fig1820residuals} and Figure \ref{FigV404residuals}. The 1D astrometric fit solutions for MAXI J1820+070 obtained perpendicular  to its jet axis (assuming a jet axis of $26^{\circ}$ in the plane of the sky) are shown in Figure \ref{fig1820results}. 
The apparent mild scatter in $5$\,GHz residuals along the jet axis (Figure \ref{Fig1820residuals}) vanishes if we ignore the singular residual data point that resides in the bottom-right quadrant of the residual plot. Our estimated parallax and proper motion from the 1D astrometry fit (provided in Table \ref{tab:results}) of MAXI J1820+070 are in strong agreement with \cite{2020MNRAS.493L..81A}, and we conclude that the previous radio astrometry was unaffected by hard state jet motion. Note that although we use the same observations of MAXI J1820+070 as \cite{2020MNRAS.493L..81A}, our 2D astrometry fit produces marginally different parallax and proper motion solutions likely due to having used different priors (we use flat priors while \cite{2020MNRAS.493L..81A} used Gaia astrometry as priors).\\

As described in Section \ref{sec1D}, our astrometry technique also fits for core shift between the different frequencies, due to contributions from the phase reference source (as shown in Figure \ref{Fig1coreshift}) and the BHXRB. \cite{2020MNRAS.493L..81A} found no evidence of core shift contribution from the phase reference source, hence using the core shift fit by our astrometry modeling we place upper limits on MAXI J1820+070's hard state jet size measured between the $5$\,GHz and $15$\,GHz photospheres to be $<0.27 $\,au measured on the plane of the sky. \cite{2020MNRAS.493L..81A} found the inclination of the jet to be $(63\pm3)^{\circ}$ using the ratio of the proper motion of approaching and receding components of the jet, and is in agreement with the values measured by \cite{2021MNRAS.505.3393W}. Assuming the hard state inclination angle to be the same as the one found by \cite{2020MNRAS.493L..81A} and \cite{2021MNRAS.505.3393W} for the outburst, we obtain an upper limit of  $0.31$ au on the displacement between the $5$\,GHz and $15$\,GHz emission surfaces measured along the jet axis.\\

Of the two BHXRBs considered, we better sample the parallax and proper motion for V404 Cygni due to how the observation epochs were spaced. While the $19$ epochs of V404 Cygni together gave us an $8$-year long time baseline to accurately constrain the proper motion, the twelve $4.8$\,GHz observations closely sampled the parallax of the source over the course of $6$ months. In Figure \ref{figV404results} we show our astrometric fitting solutions for V404 Cygni. While the $8.4$\,GHz residuals (for 2D or 1D fit) did not show any correlation with the jet axis, the $4.8$\,GHz residuals of the 1D fit showed strong scatter along the jet axis (1D fit residuals looked identical to the 2D $4.8$\,GHz residuals shown in Figure \ref{FigV404residuals} and hence are not shown again as a new figure in this paper). We also note from Figure \ref{figV404results} that the parallax of V404 Cygni estimated from the 2D and 1D fit are in slight tension with the previous VLBI and Gaia parallaxes which is further discussed below. \\

In order to verify the reliability of the $4.8$\,GHz residual scatter and the parallax discrepancy, we applied the V404 Cygni calibration solutions on the check source and imaged it using {\tt AIPS}. In Figure \ref{FigV404VsChecksource}, we show the offset of the check source (measured with respect to its median position during the $12$ epochs of $4.8$\,GHz observation) plotted against the V404 Cygni residuals along the RA and Dec axis. The two residuals seem to follow a one to one correlation, suggesting the scatter to be systematic in nature instead of being jet motion. We mitigate the systematic by re-performing our astrometry for V404 Cygni (Steps 1-4 mentioned in Section \ref{FigV404VsChecksource}), but this time with the observed check source offset subtracted from our $4.8$\,GHz target observations, and the corresponding corrected parallax and proper motion are also shown in Figure \ref{figV404results} (using purple and brown markers). The updated systematic corrected parallax measurement for V404 Cygni is now in agreement with the previous radio parallax, thus showing the previous radio astrometry by \cite{2009ApJ...706L.230M} to be unaffected by any potential jet motion. We further verify our astrometry by just using $8.4$\,GHz observations (as they had geodetic blocks for many observations and had better uv-coverage due to longer observations) and we obtain parallax measurements that are consistent (shown in Figure \ref{figV404results}) with the 1D parallax determined using both the bands, thus building confidence in our removal of systematics in the $4.8$\,GHz observations. The parallax and proper motion of the source measured along the $\theta_{\perp}$ axis is also provided in Table \ref{tab:results}. Due to an $8$ year long time baseline our uncertainties in  proper motion are also smaller (a factor of 5 along RA and a factor of 3 along DEC) than previous VLBI or Gaia measurements.\\

The astrometic fitting for V404 Cygni provided us with a core shift of $\Delta \delta = -0.19\pm0.04 $ (mas) and $\Delta \alpha = 0.38\pm0.05$ (mas) between the $8.4$\,GHz and $4.8$\,GHz data. As we are unable to isolate the core shift of V404 Cygni from the core shift of the phase reference source (according to AstroGeo\footnote{\url{http://astrogeo.org/}} images, the phase reference source had a jet extension that is fairly well aligned with the jet axis of V404 Cygni, so it is not possible to disentangle the two), we place upper limits on the jet size (measured between the $8.4$\,GHz and $4.8$\,GHz photospheres) to be $0.9^{+1.0}_{-0.8}$au (projected on the plane of the sky).  Using $67^{\circ}$\citep{2010ApJ...716.1105K} to be the jet inclination of V404 Cygni, the upper limit on jet size measured along the jet axis is $1.0^{+1.1}_{-0.9}$\,au. 


\begin{figure*}
\begin{center}
\includegraphics[width=1\linewidth,keepaspectratio]{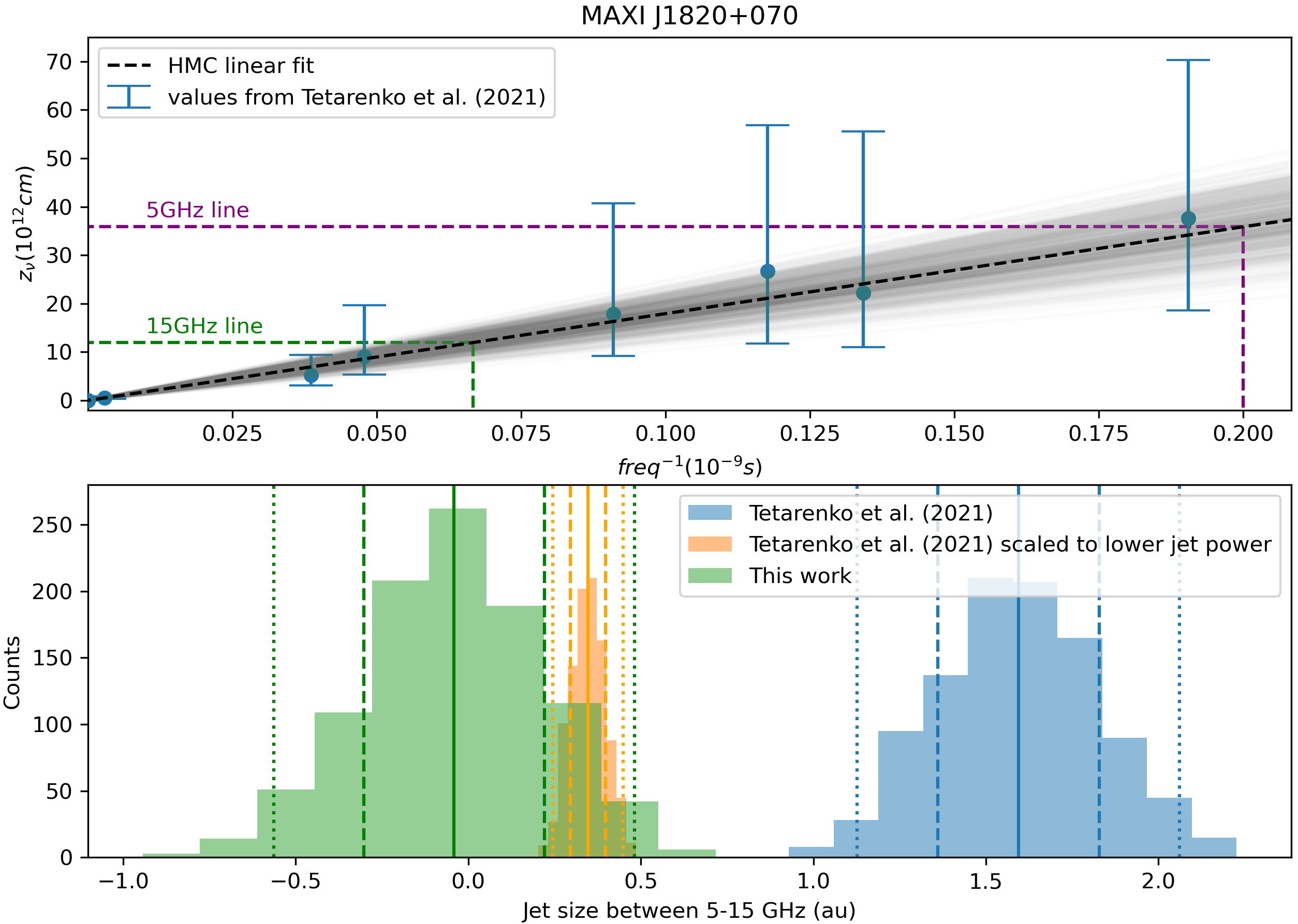}
\caption{ In the top panel we show our linear HMC fit to the measurements of the distance from the black hole (in MAXI J1820+070) to the optical depth unity surface ($Z_{\nu}$ as a function of $\frac{1}{\nu}$) from Tetarenko et al. (2021). We also show $5$\,GHz and $15$\,GHz lines between which we measure the jet scale of MAXI J1820+070 in this work. In the bottom panel, we show the jet scale between $5$\,GHz and $15$\,GHz measured using our fit to the Tetarenko et al. (2021) data in the blue histogram, and our upper limit on MAXI J1820+070's jet scale determined from our astrometry modeling in the green histogram. In orange, we show the Tetarenko et al. (2021) jet size scaled down to the radio luminosities considered in this work. For all three distributions in the bottom panel, we show the mean value using solid vertical lines, one-sigma values using dashed vertical lines, and two-sigma errors using dotted vertical lines. }
\label{fig1820Jet}
\end{center}
\end{figure*}

\begin{table*}
    \caption{The updated astrometry for V404 Cygni and MAXI J1820+070 measured perpendicular to the jet axis along with previously measured values from the literature. Note that the jet size and core shift of V404 Cygni is much larger as core shift contributions from the phase reference source have not been subtracted, and the below table does not necessarily imply a larger jet in V404 Cygni compared to MAXI J1820+070. The table also provides the lower and upper limits of the flat priors used in our astrometry modeling. }
    \centering
    \begin{tabular}{@{}cccc@{}}
    
    Parameter & flat prior used & V404 Cygni 1D astrometry results & MAXI J1820+070 1D astrometry results \\
    & & [prev. results from \cite{2009ApJ...706L.230M}] &  [prev. results from \cite{2020MNRAS.493L..81A}]\\
          \hline \hline
        $\pi$ (mas) & 0.1 to 0.9 & $0.450\pm$ 0.018 & 0.348$\pm$0.028\\
        & & [$0.418 \pm 0.024$] & [$0.348 \pm 0.033$] \\
        \hline
        $\mu_{\alpha}cos\delta$ (mas yr$^{-1}$) & -20 to 20  &-5.045$\pm$0.004 & -3.041$\pm$0.041\\
        & & [$-5.04 \pm 0.02$] & [$-3.051 \pm 0.046$]\\
        \hline
        $\mu_{\delta}$ (mas yr$^{-1}$) & -20 to 20 &-7.743$\pm$0.009 & -6.456$\pm$0.073\\
        & & [$-7.64 \pm 0.03$] & [$-6.394 \pm 0.075$]\\
        \hline
        $\Delta \alpha$(mas) & $-3.6\times 10^{6}$ to $3.6\times 10^{6}$ &-0.19$\pm$0.04 & -0.03$\pm$0.04\\
        \hline
        $\Delta \delta$(mas)& $-3.6\times 10^{6}$ to $3.6\times 10^{6}$ &0.38$\pm$0.05 & 0.03$\pm$0.08 \\
        \hline
        jet size upper limit (au) [sky plane]& NA & $0.9^{1.0}_{0.8}$  & <0.27\\
        \hline
        jet size upper limit (au) [along jet]& NA & $1.0^{1.1}_{0.9}$ & <0.31 \\
        \hline \hline
    \end{tabular}
    \label{tab:results}
\end{table*}

\section{Discussion}
\label{sec:discussion}

\subsection{MAXI J1820+070}
\label{subsec:MAXIJdiscussion}

Our 1D astrometry solutions for MAXI J1820+070 are in agreement with the previous VLBI astrometry by \cite{2020MNRAS.493L..81A}, and we find no scatter of astrometric residuals along its hard state jet axis. Using the core shift fit by our astrometry model, we place an upper limit of $0.31$\,au (1 sigma confidence) as the jet size measured between the $15$\,GHz and $5$\,GHz photosphere of MAXI J1820+070's compact jet.  The emission from the steady hard state jet is often modeled to have a conical geometry with longer wavelengths being emitted further downstream along the jet \citep{1979ApJ...232...34B}, due to the jet reaching optical depth $\tau \sim 1$ at different distances ($z_{\nu}$) from the base of the jet. Hence, emissions at longer wavelengths are delayed versions of the higher frequency emission emitted closer to the black hole. \cite{2021MNRAS.504.3862T} measured the distances of these emission regions in the high-hard state jet of MAXI J1820+070 by performing delay correlation of light curves observed at different frequencies \citep[see Table 2 in][]{2021MNRAS.504.3862T}. In order to place our jet size scale upper limits in the context of the values measured by \cite{2021MNRAS.504.3862T}, we fit a linear HMC model (as shown in the top panel of Figure \ref{fig1820Jet}), $z_{\nu} = A \times \frac{1}{\nu} + B$ (assuming $k=1$ in the \cite{1979ApJ...232...34B} jet model, where A and B are constants), to their values and obtain the jet size between the $5$\,GHz and $15$\,GHz photospheres to be $1.60\pm0.23$ au, a factor of 5 larger than our upper limit of <0.31 au (bottom panel of Figure \ref{fig1820Jet}). Note that in Figure \ref{fig1820Jet}, using the known jet axis and inclination of MAXI J1820+070 we have projected the core shift as limits on the jet size scale. Hence any negative values in the bottom panel of Figure \ref{fig1820Jet} signify a core shift in the phase reference source in the opposite direction of MAXI J1820+070's jet axis. \\

\cite{2021MNRAS.504.3862T} performed their study of MAXI J1820+070 during the high-hard state with an average flux density of $46.0 \pm 0.1$\,mJy (at $5.25$\,GHz), a factor of 25 times more luminous than our $5$\,GHz VLBI  observations (average flux density of $1.80\pm0.02$\,mJy). Hence, our smaller upper limit on the jet size scale could be due to the jet scaling down to a smaller size at lower radio luminosities. Using the analytical model for hard state jets derived by \cite{2006ApJ...636..316H}, we scale down (using $L_{\nu} \propto z_{\nu}^{\frac{17}{8}}$) the jet size ($z_{\nu}$) probed by \cite{2021MNRAS.504.3862T} to the average $5$\,GHz luminosities ($L_{\nu}$) probed by this work, and obtain a jet size of $0.35\pm0.05$\,au (one sigma errors) that is within two sigma agreement with our upper limit (also shown as the orange distribution in the bottom panel of Figure \ref{fig1820Jet})\footnote{A more recent study by \cite{2022ApJ...925..189Z} infer two different possible values for $z$ at $15$\, Ghz using two different methods ($2.5\times10^{13}$\,cm or $4\times10^{13}$\,cm). While $z=2.5\times10^{13}$\,cm provides a jet size (between $4.8 - 8.4$\, Ghz emission regions) that is within our measured upper limit (when scaled down to lower jet power), the later solution provides a jet size that is slightly discrepant with our upper limit and would only be in agreement within 3 sigma measurement errors. The later solution, if correct, would either suggest a residual core shift in the phase reference source, or require a re-examination of the jet model assumptions.}.\\

 The model of \cite{2006ApJ...636..316H} was primarily developed to study Cygnus X-1, but the equation used here to scale down the jet size was derived for a generic hard state jet. The study derives an expression for the kinetic power of the hard state jet in terms of the particle content of the jet, the filling factor, and the equipartition fraction of the magnetic field, whose values are unknown for MAXI J1820+070.\\

 \cite{2011A&A...532A..38S} found the median core shift in AGNs measured between $5$\,GHz and $15.4$\,GHz to be $0.24$ mas and hence it could be argued that the core shift fit to our MAXI J1820+070's astrometry could be affected by the core shift in the phase reference source. However, we find this to be unlikely due to the following reasons. First, even if MAXI J1820+070's phase reference sources (J1821+0549 and J1813+0615) did have a core shift large enough to impact our astrometry modelling, our upper limit on MAXI J1820+070's jet size would still be valid as both the phase reference sources show jet extensions along position angles that would add constructively to the total core shift.  Using images of the two phase-reference sources obtained from Astrogeo, we determine the core shift direction in the two systems by performing {\tt JMFIT} on the images. We obtain an approx. position angle of $9.9^{\circ +2.4}_{-2.1}$  for J1821+0549 and $85.3^{\circ +4.2}_{-5.8}$ for J1813+0615. \\
 
 Upon inspecting multi-band images of phase reference sources from Astrogeo, we find J1821+0549 to show jet extension $10^{\circ}$ west of MAXI J1820+070's jet axis, and J1813+0615 shows extension about $60^{\circ}$ further east of MAXI J1820+070's jet axis. In either scenario, it is unlikely the core shift from the phase reference source would cancel the core shift from the target source, thus demonstrating the validity of our upper limit on MAXI J1820+070's jet size. \\
 
 Second, the astrometry modelling of the source performed using multi-frequency observations in this work and \cite{2020MNRAS.493L..81A} is able to measure a parallax as small as $0.348\pm 0.033$ mas in MAXI J1820+070 with high confidence, and hence it is unlikely that our data are affected by core shifts as large as the ones reported by \cite{2011A&A...532A..38S}. \\

\subsection{V404 Cygni}
\label{subsec:V404discussion}
 

Our initial astrometric fit for V404 Cygni showed the $4.8$\,GHz observations to be affected by a systematic offset that had a one-to-one correlation with the observed offset in the check source. Upon subtracting the systematic offset and re-performing the astrometry, we obtained a parallax distance measurement of the source that is in agreement with the previous study \citep{2009ApJ...706L.230M}. We also re-performed our astrometry just using the $8.4$\,GHz observations (as they were less affected by systematics) and again obtained values that were consistent with previous studies. The parallax error bars in V404 Cygni's 1D astrometry appear slightly smaller than the error bars in the 2D astrometry, despite having lost a degree of freedom in the 1D fit, possibly due to a very mild jet scatter affecting the 2D astrometry. This could also be due to having used the posteriors of the 2D astrometry fit as priors in the 1D fit (as described in Section \ref{sec1D}). Our updated radio parallax measurement of V404 Cygni is in slight disagreement with the Gaia parallax of the system. Gaia measured the parallax of the source using regular astrometry observations performed between 2014-2017. Hence, it is possible that the optical and radio parallax discrepancy is due to the Gaia measurements potentially being affected by the 2015 outburst of V404 Cygni. \\

Using delay correlations between the hard state X-ray and radio light curves of V404 Cygni during the low-hard state, \cite{plotkin20172015} inferred the $8.4$\,GHz jet size to be $<3.0 \pm 0.8$ au. We place further limits on V404 Cygni's low-hard state jet geometry with an upper limit of $1.0\pm0.1$\,au for the jet size measured between the $8.4$\,GHz and $4.8$\,GHz photospheres. Due to the phase-reference source showing jet extensions along the jet axis of V404 Cygni, we are unable to disentangle the two. As both the BHXRBs considered in this work are at a similar distance and have similar inclination angles, assuming V404 Cygni to have a larger jet would be in agreement with the fact that V404 Cygni has a much more luminous jet, and hence the detection of a larger jet would be expected from the jet size scaling relationship derived by \cite{2006ApJ...636..316H}. However, we refrain from interpreting our core shift to be conclusive evidence of a larger jet in V404 Cygni due to, first, not having been able to remove/quantify the core shift contribution from the phase reference source, and second, due to having identified our $4.8$\,GHz observations of V404 Cygni to being affected by systematics.  Hence we only use our core shift as an upper limit on the jet size.  \\

\section{Conclusion}
\label{sec:conclustion}
In this paper, we probe the jet size scale in the hard state jets of MAXI J1820+070 and V404 Cygni whilst simultaneously investigating the possibility of previous VLBI astrometry being affected by hard state jet motion. We place an upper limit of $<0.31$\,au on the jet size measured between the 15GHz and 5GHz emission regions of the hard state jet in MAXI J1820+070. Our limit on the jet size is a factor of $5$ smaller than the size probed by \cite{2021MNRAS.504.3862T} (in the high hard state), thus showing evidence for the jet size to scale with jet luminosity. For V404 Cygni, we place an upper limit of $<1.0$\,au on the distance between $8.4-4.8$\,GHz emission regions. \\

Using a Bayesian framework, we also fit for astrometry of the two systems using 1D source positions measured perpendicular to the jet axis, which are unaffected by any potential jet motion \citep{2021Sci...371.1046M}. Our results are in strong agreement with the previous VLBI astrometry (\cite{2020MNRAS.493L..81A} for MAXI J1820+070 and \cite{2009ApJ...706L.230M} for V404 Cygni), and demonstrate the previous radio parallax measurements to be unaffected by any hard state jet motion. Due to having used fourteen new observations along with previously published epochs to perform astrometry of V404 Cygni, our updated parallax measurement of V404 Cygni is $0.450 \pm 0.0018$\,mas ($2.226 \pm 0.091 $\,kpc).\\

\section*{Acknowledgements}
This work has made use of data from the European Space Agency (ESA) mission
{\it Gaia} (\url{https://www.cosmos.esa.int/gaia}), processed by the {\it Gaia}
Data Processing and Analysis Consortium (DPAC,
\url{https://www.cosmos.esa.int/web/gaia/dpac/consortium}). Funding for the DPAC
has been provided by national institutions, in particular the institutions
participating in the {\it Gaia} Multilateral Agreement. CMW acknowledges financial support from the Forrest Research Foundation Scholarship, the Jean-Pierre Macquart Scholarship, and the Australian Government Research Training Program Scholarship. JS acknowledges support by the Packard Foundation.

\subsection*{Sofware}
We acknowledge the work and the support of the developers of the following Python packages:
Astropy \citep{theastropycollaboration_astropy_2013,astropycollaboration_astropy_2018}, Numpy \citep{vanderwalt_numpy_2011}, Scipy  \citep{jones_scipy_2001}, matplotlib \citep{Hunter:2007}, SkyField\footnote{\url{https://rhodesmill.org/skyfield/}}, and PYMC3 \citep{salvatier2016probabilistic}. 

\section*{Data Availability}
The code and data required to reproduce all the figures in this paper can be obtained from \url{https://github.com/BHXRBs/VLBI-GAIA-Astrometry-V404Cygni-MAXIJ1820}


\begin{table*}
	\centering
	\caption{MAXI J1820+070 Astrometry. The below values are neither corrected for differences in the absolute reference frame nor the core shift between the two frequencies.}
	\label{tab:J1820}
	\begin{tabular}{lccccccr} 
		\hline
		Project & Epoch  & MJD & Frequency &  Bandwidth & R.A. (J2000) & Dec. (J2000) & comment\\
        ID &  &  & (GHz) & (MHz) & (18$^h$20$^m$) & (07$^{\circ}$11$^{'}$) &\\

		\hline \hline
		BM467A & M01& 58193.65 & 15 & 32 & 21$^{s}$.9386536(1)  & 07$^{''}$.00170025(4)  & obtained from \cite{2020MNRAS.493L..81A}\\
		BM467O & M02& 58397.01 & 15 & 32 & 21$^{s}$.9384875(4)  & 07$^{''}$.00166302(10) & obtained from \cite{2020MNRAS.493L..81A}\\
		EA062A & M03& 58407.71 & 5  & 32 & 21$^{s}$.9384883(33) & 07$^{''}$.00166075(27) & obtained from \cite{2020MNRAS.493L..81A}\\
		BM467R & M04& 58441.73 & 15 & 32 & 21$^{s}$.9384770(9)  & 07$^{''}$.00165549(31) & obtained from \cite{2020MNRAS.493L..81A}\\
		EA062B & M05& 58457.04 & 5  & 32 & 21$^{s}$.938437(16)  & 07$^{''}$.0016485(12)  & obtained from \cite{2020MNRAS.493L..81A}\\
		BM467S & M06& 58474.86 & 5  & 32 & 21$^{s}$.938462(14)  & 07$^{''}$.0016498(41)  & obtained from \cite{2020MNRAS.493L..81A}\\
		EA062C & M07& 58562.25 & 5  & 32 & 21$^{s}$.9384324(12) & 07$^{''}$.00163533(10) & obtained from \cite{2020MNRAS.493L..81A}\\
		BA130B & M08& 58718.06 & 5  & 32 & 21$^{s}$.9382958(8)  & 07$^{''}$.00160872(21) & obtained from \cite{2020MNRAS.493L..81A}\\
		       & M09& 58718.14 & 15 & 32 & 21$^{s}$.9383011(3)  & 07$^{''}$.00160709(14) & obtained from \cite{2020MNRAS.493L..81A}\\
		BA130C & M10& 58755.04 & 5  & 32 & 21$^{s}$.9382761(28) & 07$^{''}$.00159845(93) & obtained from \cite{2020MNRAS.493L..81A}\\
		       & M11& 58755.12 & 15 & 32 & 21$^{s}$.9382730(22) & 07$^{''}$.00160090(74) & obtained from \cite{2020MNRAS.493L..81A}\\
		\hline \hline
	\end{tabular}
\end{table*}

\begin{table*}
	\centering
	\caption{V404 Astrometry. The below values are not corrected for core shift between the two frequencies. }
	\label{tab:v404}
	\begin{tabular}{lccccccr} 
		\hline
		Project & Epoch  & MJD & Frequency &  Bandwidth & R.A. & Dec.& comment\\
        ID &  &  & (GHz) & (MHz) & (20$^{h}4^{m}$) & (33$^{\circ}$52$^{'}$) & \\
        \hline \hline
        BG168 & V01 & 54436.84 & 8.421 & 32 & 03.821266(5) & 01.89861(22) & obtained from \citep{2009ApJ...706L.230M}\\
        BM290 & V02 & 54787.98 & 8.408 & 64 & 03.820888(3) & 01.89138(9)  & obtained from \citep{2009ApJ...706L.230M}\\
        BM290 & V03 & 54877.76 & 8.408 & 64 & 03.820826(5) & 01.88959(12) & obtained from \citep{2009ApJ...706L.230M}\\
        BM290 & V04 & 54947.57 & 8.408 & 64 & 03.820777(3) & 01.88813(8)  & obtained from \citep{2009ApJ...706L.230M}\\
        BM290 & V05 & 55015.38 & 8.408 & 64 & 03.820688(5) & 01.88741(17) & obtained from \citep{2009ApJ...706L.230M}\\
        BM290 & V06 & 55156.96 & 8.408 & 64 & 03.820496(8) & 01.88338(32) & reduced in this work\\
        BM399 & V07 & 56705.82 & 4.85 & 256 & 03.81880(1)  & 01.8510(2)   & reduced in this work\\
        BM399 & V08 & 56745.73 & 4.85 & 256 & 03.81884(2)  & 01.8493(2)   & reduced in this work\\
        BM399 & V09 & 56761.50 & 4.85 & 256 & 03.818800(9) & 01.8490(2)   & reduced in this work\\
        BM399 & V10 & 56775.63 & 4.85 & 256 & 03.818766(4) & 01.8489(1)   & reduced in this work\\
        BM399 & V11 & 56789.44 & 4.85 & 256 & 03.818750(8) & 01.8485(2)   & reduced in this work\\
        BM399 & V12 & 56805.46 & 4.85 & 256 & 03.818718(3) & 01.8485(1)   & reduced in this work\\
        BM399 & V13 & 56824.49 & 4.85 & 256 & 03.818676(7) & 01.8488(1)   & reduced in this work\\
        BM399 & V14 & 56841.45 & 4.85 & 256 & 03.818671(9) & 01.8485(2)   & reduced in this work\\
        BM399 & V15 & 56851.42 & 4.85 & 256 & 03.818633(8) & 01.8483(2)   & reduced in this work\\
        BM399 & V16 & 56862.33 & 4.85 & 256 & 03.818632(6) & 01.8480(1)   & reduced in this work\\
        BM399 & V17 & 56882.37 & 4.85 & 256 & 03.818562(2) & 01.8470(3)   & reduced in this work\\
        BM399 & V18 & 56891.38 & 4.85 & 256 & 03.818533(1) & 01.8469(1)   & reduced in this work\\
        BM421 & V19 & 57211.46 & 8.4 & 256 & 03.818241(2) & 01.84058(3)  & reduced in this work\\
        
		\hline \hline
	\end{tabular}
\end{table*}



\bibliographystyle{mnras}
\bibliography{example} 

\begin{thebibliography}{}
\makeatletter
\relax
\def\mn@urlcharsother{\let\do\@makeother \do\$\do\&\do\#\do\^\do\_\do\%\do\~}
\def\mn@doi{\begingroup\mn@urlcharsother \@ifnextchar [ {\mn@doi@}
  {\mn@doi@[]}}
\def\mn@doi@[#1]#2{\def\@tempa{#1}\ifx\@tempa\@empty \href
  {http://dx.doi.org/#2} {doi:#2}\else \href {http://dx.doi.org/#2} {#1}\fi
  \endgroup}
\def\mn@eprint#1#2{\mn@eprint@#1:#2::\@nil}
\def\mn@eprint@arXiv#1{\href {http://arxiv.org/abs/#1} {{\tt arXiv:#1}}}
\def\mn@eprint@dblp#1{\href {http://dblp.uni-trier.de/rec/bibtex/#1.xml}
  {dblp:#1}}
\def\mn@eprint@#1:#2:#3:#4\@nil{\def\@tempa {#1}\def\@tempb {#2}\def\@tempc
  {#3}\ifx \@tempc \@empty \let \@tempc \@tempb \let \@tempb \@tempa \fi \ifx
  \@tempb \@empty \def\@tempb {arXiv}\fi \@ifundefined
  {mn@eprint@\@tempb}{\@tempb:\@tempc}{\expandafter \expandafter \csname
  mn@eprint@\@tempb\endcsname \expandafter{\@tempc}}}

\bibitem[\protect\citeauthoryear{{Astropy Collaboration} et~al.,}{{Astropy
  Collaboration} et~al.}{2018}]{astropycollaboration_astropy_2018}
{Astropy Collaboration} et~al., 2018, \mn@doi [The Astronomical Journal]
  {10.3847/1538-3881/aabc4f}, 156, 123

\bibitem[\protect\citeauthoryear{{Atri} et~al.,}{{Atri}
  et~al.}{2020}]{2020MNRAS.493L..81A}
{Atri} P.,  et~al., 2020, \mn@doi [\mnras] {10.1093/mnrasl/slaa010}, \href
  {https://ui.adsabs.harvard.edu/abs/2020MNRAS.493L..81A} {493, L81}

\bibitem[\protect\citeauthoryear{{Blandford} \& {K{\"o}nigl}}{{Blandford} \&
  {K{\"o}nigl}}{1979}]{1979ApJ...232...34B}
{Blandford} R.~D.,  {K{\"o}nigl} A.,  1979, \mn@doi [\apj] {10.1086/157262},
  \href {https://ui.adsabs.harvard.edu/abs/1979ApJ...232...34B} {232, 34}

\bibitem[\protect\citeauthoryear{{Brandt}}{{Brandt}}{2018}]{2018ApJS..239...31B}
{Brandt} T.~D.,  2018, \mn@doi [\apjs] {10.3847/1538-4365/aaec06}, \href
  {https://ui.adsabs.harvard.edu/abs/2018ApJS..239...31B} {239, 31}

\bibitem[\protect\citeauthoryear{{Bright} et~al.,}{{Bright}
  et~al.}{2020}]{2020NatAs...4..697B}
{Bright} J.~S.,  et~al., 2020, \mn@doi [Nature Astronomy]
  {10.1038/s41550-020-1023-5}, \href
  {https://ui.adsabs.harvard.edu/abs/2020NatAs...4..697B} {4, 697}

\bibitem[\protect\citeauthoryear{{Casares}, {Charles}  \& {Naylor}}{{Casares}
  et~al.}{1992}]{1992Natur.355..614C}
{Casares} J.,  {Charles} P.~A.,   {Naylor} T.,  1992, \mn@doi [\nat]
  {10.1038/355614a0}, \href
  {https://ui.adsabs.harvard.edu/abs/1992Natur.355..614C} {355, 614}

\bibitem[\protect\citeauthoryear{{Corral-Santana}, {Casares},
  {Mu{\~n}oz-Darias}, {Bauer}, {Mart{\'\i}nez-Pais}  \&
  {Russell}}{{Corral-Santana} et~al.}{2016}]{2016A&A...587A..61C}
{Corral-Santana} J.~M.,  {Casares} J.,  {Mu{\~n}oz-Darias} T.,  {Bauer} F.~E.,
  {Mart{\'\i}nez-Pais} I.~G.,   {Russell} D.~M.,  2016, \mn@doi [\aap]
  {10.1051/0004-6361/201527130}, \href
  {https://ui.adsabs.harvard.edu/abs/2016A&A...587A..61C} {587, A61}

\bibitem[\protect\citeauthoryear{{Dhawan}, {Mirabel}  \&
  {Rodr{\'\i}guez}}{{Dhawan} et~al.}{2000}]{2000ApJ...543..373D}
{Dhawan} V.,  {Mirabel} I.~F.,   {Rodr{\'\i}guez} L.~F.,  2000, \mn@doi [\apj]
  {10.1086/317088}, \href
  {https://ui.adsabs.harvard.edu/abs/2000ApJ...543..373D} {543, 373}

\bibitem[\protect\citeauthoryear{Dixon}{Dixon}{2006}]{dixon2006bootstrap}
Dixon P.~M.,  2006, Encyclopedia of environmetrics

\bibitem[\protect\citeauthoryear{{Espinasse} et~al.,}{{Espinasse}
  et~al.}{2020}]{2020ApJ...895L..31E}
{Espinasse} M.,  et~al., 2020, \mn@doi [\apjl] {10.3847/2041-8213/ab88b6},
  \href {https://ui.adsabs.harvard.edu/abs/2020ApJ...895L..31E} {895, L31}

\bibitem[\protect\citeauthoryear{{Fender}}{{Fender}}{2006}]{2006csxs.book..381F}
{Fender} R.,  2006, in , Vol.~39, Compact stellar X-ray sources.
pp 381--419, \mn@doi{10.48550/arXiv.astro-ph/0303339}

\bibitem[\protect\citeauthoryear{{Fender} \& {Gallo}}{{Fender} \&
  {Gallo}}{2014}]{2014SSRv..183..323F}
{Fender} R.,  {Gallo} E.,  2014, \mn@doi [\ssr] {10.1007/s11214-014-0069-z},
  \href {https://ui.adsabs.harvard.edu/abs/2014SSRv..183..323F} {183, 323}

\bibitem[\protect\citeauthoryear{{Fender} \& {Mu{\~n}oz-Darias}}{{Fender} \&
  {Mu{\~n}oz-Darias}}{2016}]{2016LNP...905...65F}
{Fender} R.,  {Mu{\~n}oz-Darias} T.,  2016, in {Haardt} F.,  {Gorini} V.,
  {Moschella} U.,  {Treves} A.,   {Colpi} M.,  eds, , Vol.~905, Lecture Notes
  in Physics, Berlin Springer Verlag.
Springer International Publishing, p.~65, \mn@doi{10.1007/978-3-319-19416-5_3}

\bibitem[\protect\citeauthoryear{{Fromm}, {Perucho}, {Savolainen}, {Ros},
  {Lobanov}, {Zensus}  \& {L{\"a}hteenm{\"a}ki}}{{Fromm}
  et~al.}{2011}]{2011IAUS..275..194F}
{Fromm} C.~M.,  {Perucho} M.,  {Savolainen} T.,  {Ros} E.,  {Lobanov} A.~P.,
  {Zensus} J.~A.,   {L{\"a}hteenm{\"a}ki} A.,  2011, in {Romero} G.~E.,
  {Sunyaev} R.~A.,   {Belloni} T.,  eds, ~ Vol. 275, Jets at All Scales. pp
  194--195 (\mn@eprint {arXiv} {1011.4837}), \mn@doi{10.1017/S1743921310016017}

\bibitem[\protect\citeauthoryear{{Gaia Collaboration} et~al.,}{{Gaia
  Collaboration} et~al.}{2016}]{2016A&A...595A...1G}
{Gaia Collaboration} et~al., 2016, \mn@doi [\aap]
  {10.1051/0004-6361/201629272}, \href
  {https://ui.adsabs.harvard.edu/abs/2016A&A...595A...1G} {595, A1}

\bibitem[\protect\citeauthoryear{Gelman \& Rubin}{Gelman \&
  Rubin}{1992}]{gelman1992inference}
Gelman A.,  Rubin D.~B.,  1992, Statistical science, 7, 457

\bibitem[\protect\citeauthoryear{Greisen}{Greisen}{2003}]{Greisen2003}
Greisen E.~W.,  2003, in , Information Handling in Astronomy - Historical
  Vistas.
Springer Netherlands, pp 109--125, \mn@doi{10.1007/0-306-48080-8_7}, \url
  {https://doi.org/10.1007/0-306-48080-8_7}

\bibitem[\protect\citeauthoryear{{Groenewegen}}{{Groenewegen}}{2021}]{2021A&A...654A..20G}
{Groenewegen} M.~A.~T.,  2021, \mn@doi [\aap] {10.1051/0004-6361/202140862},
  \href {https://ui.adsabs.harvard.edu/abs/2021A&A...654A..20G} {654, A20}

\bibitem[\protect\citeauthoryear{{Hada}, {Doi}, {Kino}, {Nagai}, {Hagiwara}  \&
  {Kawaguchi}}{{Hada} et~al.}{2012}]{2012Natur.489..326H}
{Hada} K.,  {Doi} A.,  {Kino} M.,  {Nagai} H.,  {Hagiwara} Y.,   {Kawaguchi}
  N.,  2012, \mn@doi [\nat] {10.1038/nature11425}, \href
  {https://ui.adsabs.harvard.edu/abs/2012Natur.489..326H} {489, 326}

\bibitem[\protect\citeauthoryear{{Han} \& {Hjellming}}{{Han} \&
  {Hjellming}}{1992}]{1992ApJ...400..304H}
{Han} X.,  {Hjellming} R.~M.,  1992, \mn@doi [\apj] {10.1086/171996}, \href
  {https://ui.adsabs.harvard.edu/abs/1992ApJ...400..304H} {400, 304}

\bibitem[\protect\citeauthoryear{{Heinz}}{{Heinz}}{2006}]{2006ApJ...636..316H}
{Heinz} S.,  2006, \mn@doi [\apj] {10.1086/497954}, \href
  {https://ui.adsabs.harvard.edu/abs/2006ApJ...636..316H} {636, 316}

\bibitem[\protect\citeauthoryear{Hunter}{Hunter}{2007}]{Hunter:2007}
Hunter J.~D.,  2007, \mn@doi [Computing in Science \& Engineering]
  {10.1109/MCSE.2007.55}, 9, 90

\bibitem[\protect\citeauthoryear{Jones, Oliphant, Peterson  \& {Others}}{Jones
  et~al.}{2001}]{jones_scipy_2001}
Jones E.,  Oliphant T.,  Peterson P.,   {Others} 2001, {{SciPy}}: {{Open}}
  Source Scientific Tools for {{Python}}

\bibitem[\protect\citeauthoryear{{Kawamuro} et~al.,}{{Kawamuro}
  et~al.}{2018}]{2018ATel11399....1K}
{Kawamuro} T.,  et~al., 2018, The Astronomer's Telegram, \href
  {https://ui.adsabs.harvard.edu/abs/2018ATel11399....1K} {11399, 1}

\bibitem[\protect\citeauthoryear{{Khargharia}, {Froning}  \&
  {Robinson}}{{Khargharia} et~al.}{2010}]{2010ApJ...716.1105K}
{Khargharia} J.,  {Froning} C.~S.,   {Robinson} E.~L.,  2010, \mn@doi [\apj]
  {10.1088/0004-637X/716/2/1105}, \href
  {https://ui.adsabs.harvard.edu/abs/2010ApJ...716.1105K} {716, 1105}

\bibitem[\protect\citeauthoryear{{Kudryavtseva}, {Gabuzda}, {Aller}  \&
  {Aller}}{{Kudryavtseva} et~al.}{2011}]{2011MNRAS.415.1631K}
{Kudryavtseva} N.~A.,  {Gabuzda} D.~C.,  {Aller} M.~F.,   {Aller} H.~D.,  2011,
  \mn@doi [\mnras] {10.1111/j.1365-2966.2011.18808.x}, \href
  {https://ui.adsabs.harvard.edu/abs/2011MNRAS.415.1631K} {415, 1631}

\bibitem[\protect\citeauthoryear{{Lindegren} et~al.,}{{Lindegren}
  et~al.}{2021}]{2021A&A...649A...4L}
{Lindegren} L.,  et~al., 2021, \mn@doi [\aap] {10.1051/0004-6361/202039653},
  \href {https://ui.adsabs.harvard.edu/abs/2021A&A...649A...4L} {649, A4}

\bibitem[\protect\citeauthoryear{{Loinard}, {Torres}, {Mioduszewski},
  {Rodr{\'\i}guez}, {Gonz{\'a}lez-L{\'o}pezlira}, {Lachaume}, {V{\'a}zquez}  \&
  {Gonz{\'a}lez}}{{Loinard} et~al.}{2007}]{2007ApJ...671..546L}
{Loinard} L.,  {Torres} R.~M.,  {Mioduszewski} A.~J.,  {Rodr{\'\i}guez} L.~F.,
  {Gonz{\'a}lez-L{\'o}pezlira} R.~A.,  {Lachaume} R.,  {V{\'a}zquez} V.,
  {Gonz{\'a}lez} E.,  2007, \mn@doi [\apj] {10.1086/522493}, \href
  {https://ui.adsabs.harvard.edu/abs/2007ApJ...671..546L} {671, 546}

\bibitem[\protect\citeauthoryear{{Miller-Jones}, {Jonker}, {Dhawan}, {Brisken},
  {Rupen}, {Nelemans}  \& {Gallo}}{{Miller-Jones}
  et~al.}{2009}]{2009ApJ...706L.230M}
{Miller-Jones} J.~C.~A.,  {Jonker} P.~G.,  {Dhawan} V.,  {Brisken} W.,  {Rupen}
  M.~P.,  {Nelemans} G.,   {Gallo} E.,  2009, \mn@doi [\apjl]
  {10.1088/0004-637X/706/2/L230}, \href
  {https://ui.adsabs.harvard.edu/abs/2009ApJ...706L.230M} {706, L230}

\bibitem[\protect\citeauthoryear{{Miller-Jones} et~al.,}{{Miller-Jones}
  et~al.}{2019}]{2019Natur.569..374M}
{Miller-Jones} J. C.~A.,  et~al., 2019, \mn@doi [\nat]
  {10.1038/s41586-019-1152-0}, \href
  {https://ui.adsabs.harvard.edu/abs/2019Natur.569..374M} {569, 374}

\bibitem[\protect\citeauthoryear{{Miller-Jones} et~al.,}{{Miller-Jones}
  et~al.}{2021}]{2021Sci...371.1046M}
{Miller-Jones} J. C.~A.,  et~al., 2021, \mn@doi [Science]
  {10.1126/science.abb3363}, \href
  {https://ui.adsabs.harvard.edu/abs/2021Sci...371.1046M} {371, 1046}

\bibitem[\protect\citeauthoryear{Mioduszewski \& Kogan}{Mioduszewski \&
  Kogan}{2009}]{mioduszewski2009strategy}
Mioduszewski A.~J.,  Kogan L.,  2009, Technical report, Strategy for Removing
  Tropospheric and Clock Errors using delzn Version 2.0.
AIPS Memo 110, National Radio Astronomy Observatory and Cornell University

\bibitem[\protect\citeauthoryear{{Neal}}{{Neal}}{2011}]{2011hmcm.book..113N}
{Neal} R.,  2011, in , Handbook of Markov Chain Monte Carlo.
Chapman and Hall/{CRC}, pp 113--162, \mn@doi{10.1201/b10905}

\bibitem[\protect\citeauthoryear{{O'Sullivan} \& {Gabuzda}}{{O'Sullivan} \&
  {Gabuzda}}{2009}]{2009MNRAS.400...26O}
{O'Sullivan} S.~P.,  {Gabuzda} D.~C.,  2009, \mn@doi [\mnras]
  {10.1111/j.1365-2966.2009.15428.x}, \href
  {https://ui.adsabs.harvard.edu/abs/2009MNRAS.400...26O} {400, 26}

\bibitem[\protect\citeauthoryear{Plotkin et~al.,}{Plotkin
  et~al.}{2017}]{plotkin20172015}
Plotkin R.,  et~al., 2017, The Astrophysical Journal, 834, 104

\bibitem[\protect\citeauthoryear{{Plotkin}, {Miller-Jones}, {Chomiuk},
  {Strader}, {Bruzewski}, {Bundas}, {Smith}  \& {Ruan}}{{Plotkin}
  et~al.}{2019}]{2019ApJ...874...13P}
{Plotkin} R.~M.,  {Miller-Jones} J.~C.~A.,  {Chomiuk} L.,  {Strader} J.,
  {Bruzewski} S.,  {Bundas} A.,  {Smith} K.~R.,   {Ruan} J.~J.,  2019, \mn@doi
  [\apj] {10.3847/1538-4357/ab01cc}, \href
  {https://ui.adsabs.harvard.edu/abs/2019ApJ...874...13P} {874, 13}

\bibitem[\protect\citeauthoryear{{Pushkarev}, {Hovatta}, {Kovalev}, {Lister},
  {Lobanov}, {Savolainen}  \& {Zensus}}{{Pushkarev}
  et~al.}{2012}]{2012A&A...545A.113P}
{Pushkarev} A.~B.,  {Hovatta} T.,  {Kovalev} Y.~Y.,  {Lister} M.~L.,  {Lobanov}
  A.~P.,  {Savolainen} T.,   {Zensus} J.~A.,  2012, \mn@doi [\aap]
  {10.1051/0004-6361/201219173}, \href
  {https://ui.adsabs.harvard.edu/abs/2012A&A...545A.113P} {545, A113}

\bibitem[\protect\citeauthoryear{{Remillard} \& {McClintock}}{{Remillard} \&
  {McClintock}}{2006}]{2006ARA&A..44...49R}
{Remillard} R.~A.,  {McClintock} J.~E.,  2006, \mn@doi [\araa]
  {10.1146/annurev.astro.44.051905.092532}, \href
  {https://ui.adsabs.harvard.edu/abs/2006ARA&A..44...49R} {44, 49}

\bibitem[\protect\citeauthoryear{{Rushton} et~al.,}{{Rushton}
  et~al.}{2012}]{2012MNRAS.419.3194R}
{Rushton} A.,  et~al., 2012, \mn@doi [\mnras]
  {10.1111/j.1365-2966.2011.19959.x}, \href
  {https://ui.adsabs.harvard.edu/abs/2012MNRAS.419.3194R} {419, 3194}

\bibitem[\protect\citeauthoryear{{Russell} et~al.,}{{Russell}
  et~al.}{2015}]{2015MNRAS.450.1745R}
{Russell} T.~D.,  et~al., 2015, \mn@doi [\mnras] {10.1093/mnras/stv723}, \href
  {https://ui.adsabs.harvard.edu/abs/2015MNRAS.450.1745R} {450, 1745}

\bibitem[\protect\citeauthoryear{Salvatier, Wiecki  \& Fonnesbeck}{Salvatier
  et~al.}{2016}]{salvatier2016probabilistic}
Salvatier J.,  Wiecki T.~V.,   Fonnesbeck C.,  2016, PeerJ Computer Science, 2,
  e55

\bibitem[\protect\citeauthoryear{{Shahbaz}, {Ringwald}, {Bunn}, {Naylor},
  {Charles}  \& {Casares}}{{Shahbaz} et~al.}{1994}]{1994MNRAS.271L..10S}
{Shahbaz} T.,  {Ringwald} F.~A.,  {Bunn} J.~C.,  {Naylor} T.,  {Charles} P.~A.,
    {Casares} J.,  1994, \mn@doi [\mnras] {10.1093/mnras/271.1.L10}, \href
  {https://ui.adsabs.harvard.edu/abs/1994MNRAS.271L..10S} {271, L10}

\bibitem[\protect\citeauthoryear{{Shepherd}}{{Shepherd}}{1997}]{1997ASPC..125...77S}
{Shepherd} M.~C.,  1997, in {Hunt} G.,  {Payne} H.,  eds,  Astronomical Society
  of the Pacific Conference Series Vol. 125, Astronomical Data Analysis
  Software and Systems VI. p.~77

\bibitem[\protect\citeauthoryear{{Sokolovsky}, {Kovalev}, {Pushkarev}  \&
  {Lobanov}}{{Sokolovsky} et~al.}{2011}]{2011A&A...532A..38S}
{Sokolovsky} K.~V.,  {Kovalev} Y.~Y.,  {Pushkarev} A.~B.,   {Lobanov} A.~P.,
  2011, \mn@doi [\aap] {10.1051/0004-6361/201016072}, \href
  {https://ui.adsabs.harvard.edu/abs/2011A&A...532A..38S} {532, A38}

\bibitem[\protect\citeauthoryear{Stirling, Spencer  \& Garrett}{Stirling
  et~al.}{1998}]{stirling1998high}
Stirling A.,  Spencer R.,   Garrett M.,  1998, New Astronomy Reviews, 42, 657

\bibitem[\protect\citeauthoryear{{Tetarenko}, {Casella}, {Miller-Jones},
  {Sivakoff}, {Tetarenko}, {Maccarone}, {Gandhi}  \& {Eikenberry}}{{Tetarenko}
  et~al.}{2019}]{2019MNRAS.484.2987T}
{Tetarenko} A.~J.,  {Casella} P.,  {Miller-Jones} J.~C.~A.,  {Sivakoff} G.~R.,
  {Tetarenko} B.~E.,  {Maccarone} T.~J.,  {Gandhi} P.,   {Eikenberry} S.,
  2019, \mn@doi [\mnras] {10.1093/mnras/stz165}, \href
  {https://ui.adsabs.harvard.edu/abs/2019MNRAS.484.2987T} {484, 2987}

\bibitem[\protect\citeauthoryear{{Tetarenko} et~al.,}{{Tetarenko}
  et~al.}{2021}]{2021MNRAS.504.3862T}
{Tetarenko} A.~J.,  et~al., 2021, \mn@doi [\mnras] {10.1093/mnras/stab820},
  \href {https://ui.adsabs.harvard.edu/abs/2021MNRAS.504.3862T} {504, 3862}

\bibitem[\protect\citeauthoryear{{The Astropy Collaboration} et~al.,}{{The
  Astropy Collaboration} et~al.}{2013}]{theastropycollaboration_astropy_2013}
{The Astropy Collaboration} et~al., 2013, \mn@doi [Astronomy \& Astrophysics]
  {10.1051/0004-6361/201322068}, 558, 9

\bibitem[\protect\citeauthoryear{{Torres}, {Casares}, {Jim{\'e}nez-Ibarra},
  {Mu{\~n}oz-Darias}, {Armas Padilla}, {Jonker}  \& {Heida}}{{Torres}
  et~al.}{2019}]{2019ApJ...882L..21T}
{Torres} M.~A.~P.,  {Casares} J.,  {Jim{\'e}nez-Ibarra} F.,  {Mu{\~n}oz-Darias}
  T.,  {Armas Padilla} M.,  {Jonker} P.~G.,   {Heida} M.,  2019, \mn@doi
  [\apjl] {10.3847/2041-8213/ab39df}, \href
  {https://ui.adsabs.harvard.edu/abs/2019ApJ...882L..21T} {882, L21}

\bibitem[\protect\citeauthoryear{{Tucker} et~al.,}{{Tucker}
  et~al.}{2018}]{2018ApJ...867L...9T}
{Tucker} M.~A.,  et~al., 2018, \mn@doi [\apjl] {10.3847/2041-8213/aae88a},
  \href {https://ui.adsabs.harvard.edu/abs/2018ApJ...867L...9T} {867, L9}

\bibitem[\protect\citeauthoryear{{Turon}}{{Turon}}{1995}]{1995ESASP.379..109T}
{Turon} C.,  1995, in {Perryman} M.~A.~C.,  {van Leeuwen} F.,  eds,  ESA
  Special Publication Vol. 379, Future Possibilities for bstrometry in Space.
  p.~109

\bibitem[\protect\citeauthoryear{Vallenari et~al.,}{Vallenari
  et~al.}{2022}]{gaiacollaboration2022gaia}
Vallenari A.,  et~al., 2022, Gaia Data Release 3: Summary of the content and
  survey properties (\mn@eprint {arXiv} {2208.00211})

\bibitem[\protect\citeauthoryear{{Wood} et~al.,}{{Wood}
  et~al.}{2021}]{2021MNRAS.505.3393W}
{Wood} C.~M.,  et~al., 2021, \mn@doi [\mnras] {10.1093/mnras/stab1479}, \href
  {https://ui.adsabs.harvard.edu/abs/2021MNRAS.505.3393W} {505, 3393}

\bibitem[\protect\citeauthoryear{{Zdziarski}, {Tetarenko}  \&
  {Sikora}}{{Zdziarski} et~al.}{2022}]{2022ApJ...925..189Z}
{Zdziarski} A.~A.,  {Tetarenko} A.~J.,   {Sikora} M.,  2022, \mn@doi [\apj]
  {10.3847/1538-4357/ac38a9}, \href
  {https://ui.adsabs.harvard.edu/abs/2022ApJ...925..189Z} {925, 189}

\bibitem[\protect\citeauthoryear{Zensus, Diamond, Napier  et~al.}{Zensus
  et~al.}{1995}]{zensus1995very}
Zensus J.~A.,  Diamond P.,  Napier P.,   et~al., 1995, Very long baseline
  interferometry and the VLBA.
~ Vol. 82, Astronomical Society of the Pacific

\bibitem[\protect\citeauthoryear{{dePolo}, {Plotkin}, {Miller-Jones},
  {Strader}, {Maccarone}, {O'Doherty}, {Chomiuk}  \& {Gallo}}{{dePolo}
  et~al.}{2022}]{2022MNRAS.516.4640D}
{dePolo} D.~L.,  {Plotkin} R.~M.,  {Miller-Jones} J. C.~A.,  {Strader} J.,
  {Maccarone} T.~J.,  {O'Doherty} T.~N.,  {Chomiuk} L.,   {Gallo} E.,  2022,
  \mn@doi [\mnras] {10.1093/mnras/stac2572}, \href
  {https://ui.adsabs.harvard.edu/abs/2022MNRAS.516.4640D} {516, 4640}

\bibitem[\protect\citeauthoryear{{van der Walt}, Colbert  \& Varoquaux}{{van
  der Walt} et~al.}{2011}]{vanderwalt_numpy_2011}
{van der Walt} S.,  Colbert S.~C.,   Varoquaux G.,  2011, \mn@doi [Computing in
  Science \& Engineering] {10.1109/MCSE.2011.37}, 13, 22

\makeatother
\end{thebibliography}

\appendix

\section{Orbital Motion of the two BHXRBs}
\label{sec:orbitalmotion}
In this section, we show the orbital motion of the black hole around the centre of mass of the binary system to be smaller than the angular scales probed by VLBI. In the below derivation, properties of the black hole are defined using subscript $1$ and for the donor star we use subscript $2$. \\

From Kepler's third law we know that,

\begin{equation}
    \begin{array}{l}
        
    GM_{T} = \omega^{2}a^{3}
    
    \end{array}
    \label{E.1}
\end{equation}

where $G$ is the Gravitational constant, $M_{T}$ is the total mass of the binary system, $\omega$ is the orbital frequency of the orbit ($\omega = 2\pi/P_{orb}$), and $a$ is the semi-major axis.\\

Using centre of mass and semi-major axis relations, we get,

\begin{equation}
    \begin{array}{l}
        
    M_{1}a_{1} = M_{2}a_{2} \\
    a = a_{1} + a_{2} = a_{1} (1 + \frac{a_{2}}{a_{1}}) = a_{1} (1 + \frac{M_{1}}{M_{2}}) \\
    \Rightarrow a = \frac{a_{1}}{M_{2}} (M_{T})
    
    \end{array}
    \label{E.2}
\end{equation}

Substituting \ref{E.2} in \ref{E.1}, 

\begin{equation}
    \begin{array}{l}
        
    a_{1} = \sqrt[3]{\frac {GM_{2}^{3}} {\omega^{2} M_{T}^{2}}}
    \end{array}
    \label{E.3}
\end{equation}

Using the dynamical parameters for V404 Cygni and MAXI J1820+070 obtained from BlackCAT \citep{2016A&A...587A..61C}, the orbital radius of the black holes in the two BHXRBs come out to be $1.4\times 10^{9}$\,m and $3.0\times 10^{8}$\,m respectively. Given the distance of the two binary systems from Earth, the BH's orbital motion would subtend an angle of $3.8\times 10^{-3}$mas for V404 Cygni and $6.7\times 10^{-4}$mas for MAXI J1820+070, and are not resolvable within the precision of the observations used here.


\bsp	
\label{lastpage}
\end{document}